\documentclass[12pt]{iopart}
\usepackage{hyperref}
\hypersetup{
  colorlinks,%
  citecolor=black,%
  filecolor=black,%
  linkcolor=black,%
  urlcolor=blue
  }
  
\usepackage{multirow}
\usepackage{float}
\usepackage[sorting=none, backend=bibtex]{biblatex}
\usepackage{wrapfig}
\usepackage{orcidlink}
\usepackage{soul}

\usepackage{lineno}

\newcommand{\orcidA}{\orcidlink{0000-0003-2849-0120}} % Add \orcidA{} behind the author's name // BG
\newcommand{\orcidB}{\orcidlink{0000-0001-9223-6480}} % Add \orcidB{} behind the author's name // BGG
\newcommand{\orcidC}{\orcidlink{0009-0006-9345-9620}} % Add \orcidC{} behind the author's name // PL

\begin{document}

\title{Dedicated Analysis Facility for HEP Experiments}

\author{Gábor Bíró$^{1,2}$\orcidA{}, Gergely Gábor Barnaföldi$^1$\orcidB{}, Péter Lévai$^1$\orcidC{}}
   
\address{$^1$Wigner Research Center for Physics, 29--33 Konkoly--Thege Mikl\'os Str., H-1121 Budapest, Hungary.}
\address{$^2$Institute of Physics, E\"otv\"os Lor\'and University, 1/A P\'azm\'any P\'eter S\'et\'any, H-1117  Budapest, Hungary.}
% \address{$^3$European Organization for Nuclear Research (CERN), Geneva, Switzerland.}

\ead{biro.gabor@wigner.hu; barnafoldi.gergely@wigner.hu; levai.peter@wigner.hu}

\begin{indented}
  \item[]\today
  \end{indented}
 
\begin{abstract}
High-energy physics (HEP) provides ever-growing amount of data. To analyse these, continuously-evolving computational power is required in parallel by extending the storage capacity. Such developments play key roles in the future of this field however, these can be achieved also by optimization of existing IT resources. 

One of the main computing capacity consumers in the HEP software workflow are detector simulation and data analysis. To optimize the resource requirements for these aims, the concept of a dedicated Analysis Facility (AF) for Run 3 has been suggested by the ALICE experiment at CERN. These AFs are special computing centres with a combination of CPU and fast interconnected disk storage modules, allowing for rapid turnaround of analysis tasks on a dedicated subset of data. This in turn allows for optimization of the analysis process and the codes before the analysis is performed on the large data samples on the Worldwide LHC Computing Grid.

In this paper, the structure and the progress summary of the Wigner Analysis Facility (Wigner AF) is presented for the period 2020-2022.
\end{abstract}

%\maketitle

%-------------------------------------------------------------------

\section{Introduction}
\label{sec:introduction}

The largest detectors of the Large Hadron Collider (LHC) underwent major upgrades during the Long Shutdown 2 (LS2) in the period, 2019-2022~\cite{Evans:2008zzb}. Detector sensitivity, readout hardware, indeed the associated online and offline softwares were replaced and modernized. The goal of the R\&D activities was to enable the experiments to pursue new physics in the Run-3 data taking period (2022-2025) and beyond. 

For these aims, efficient data processing was investigated on large data samples from LHC's Run 1 and Run 2. The performance of the Monte Carlo simulations were also tested and optimized for massively parallel event generation on the Worldwide LHC Computing Grid (WLCG). Finally, the aim is to achieve the best computing performance beside keeping the maintenance and operation costs at a reasonable level -- despite the age of the existing hardware components.  

The Analysis Facility at the Wigner Datacenter (WDC) were established alongside the original structure, inherited by the former CERN's Tier 0 site (Budapest, Hungary)~\cite{ALICE-PUBLIC-2021-007}. Based on the existing hardware, the topology and the modules were further optimized according to the needs of rapid data campaigns of the experimental requirements. The original idea of the offline Analysis Facility (AF)~\cite{GSIAF} was applied first in large scale at the Wigner AF, where the recent multi-core analysis software framework Hyperloop~\cite{hyperloop} was also tested. 

%--------------------------------------------------------------------
\section{Structure of the Analysis Facility}
\label{sec:structure}

The Wigner Analysis Facility is the part of the Wigner Scientific Computing Laboratory (WSCLAB), located physically in the WDC. The majority of the Wigner AF's hardware is built from the legacy hardware of the Budapest Tier 0 computing center, mostly AMD Opteron 6276 CPUs~\cite{amdopteron}.  The main purpose of the analysis facility is to efficiently process a considerable amount, $\mathcal{O}($PB$)$ of data on a daily basis while being able to scale up the resources, $\mathcal{O}(15\%)$ per year. For this reason, it is essential to have a modular design for both the storage and compute parts, and to ensure high bandwidth communication between them.

\begin{figure}[h]
\centering
\includegraphics[height=0.35\textheight]{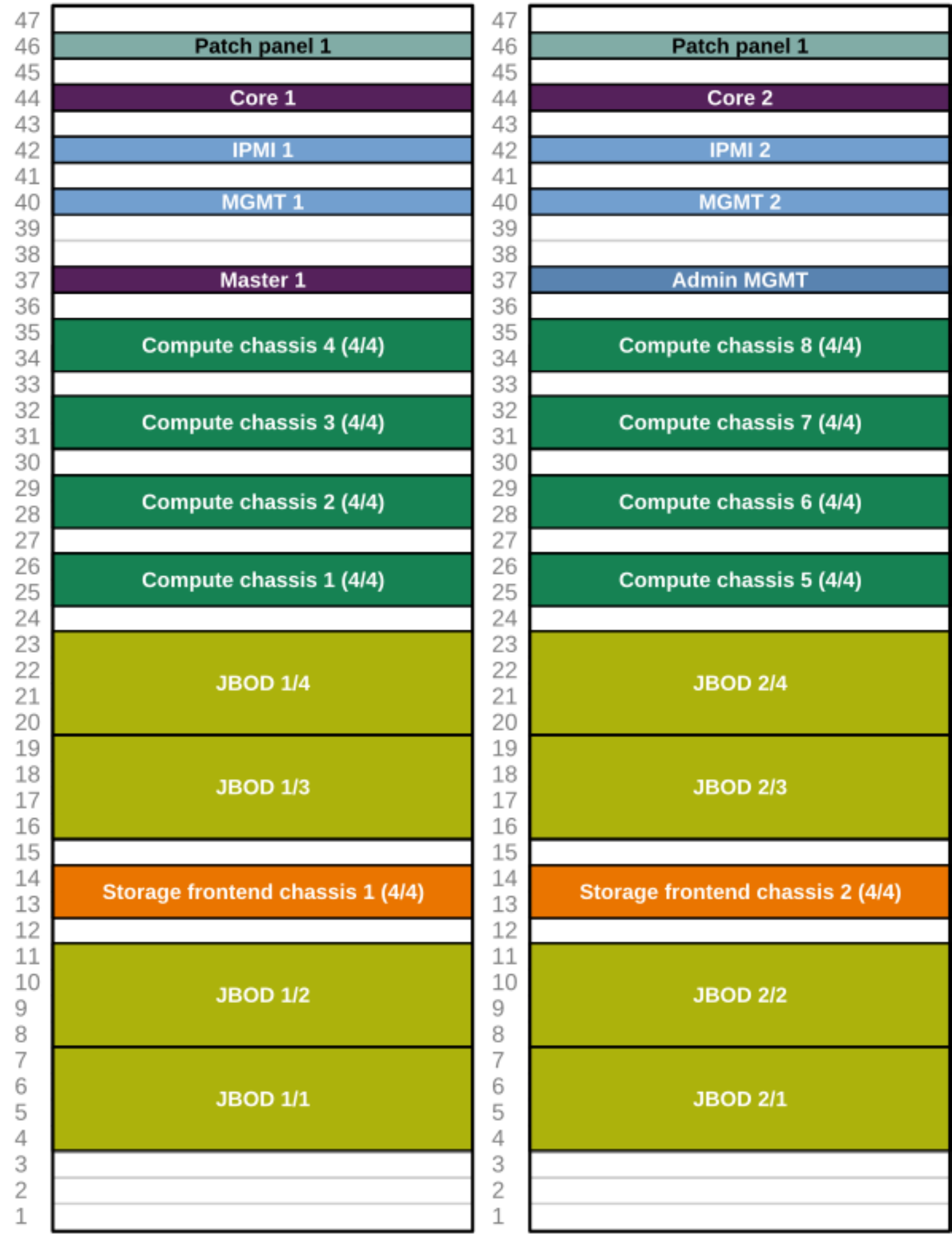}
\hfill
\includegraphics[height=0.348\textheight]{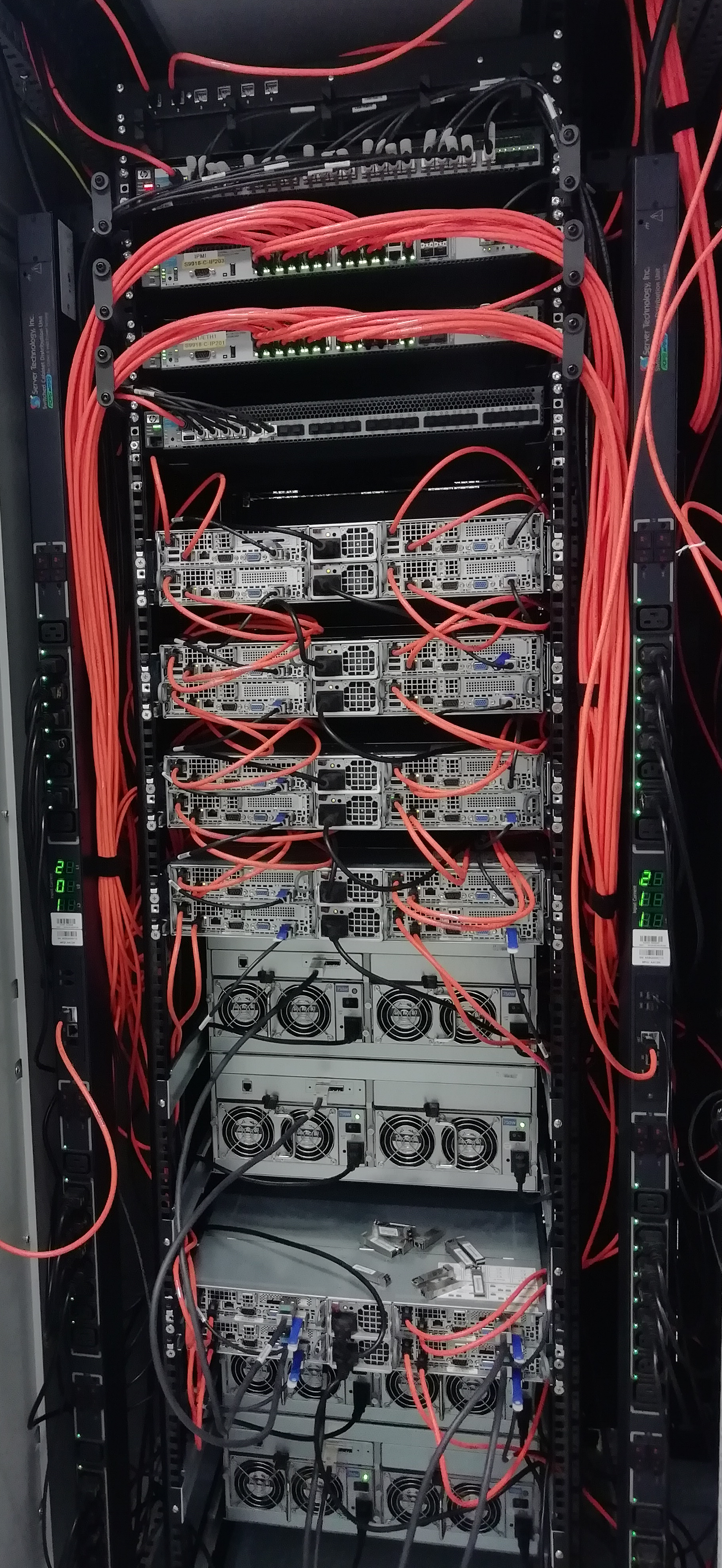}
\includegraphics[height=0.348\textheight]{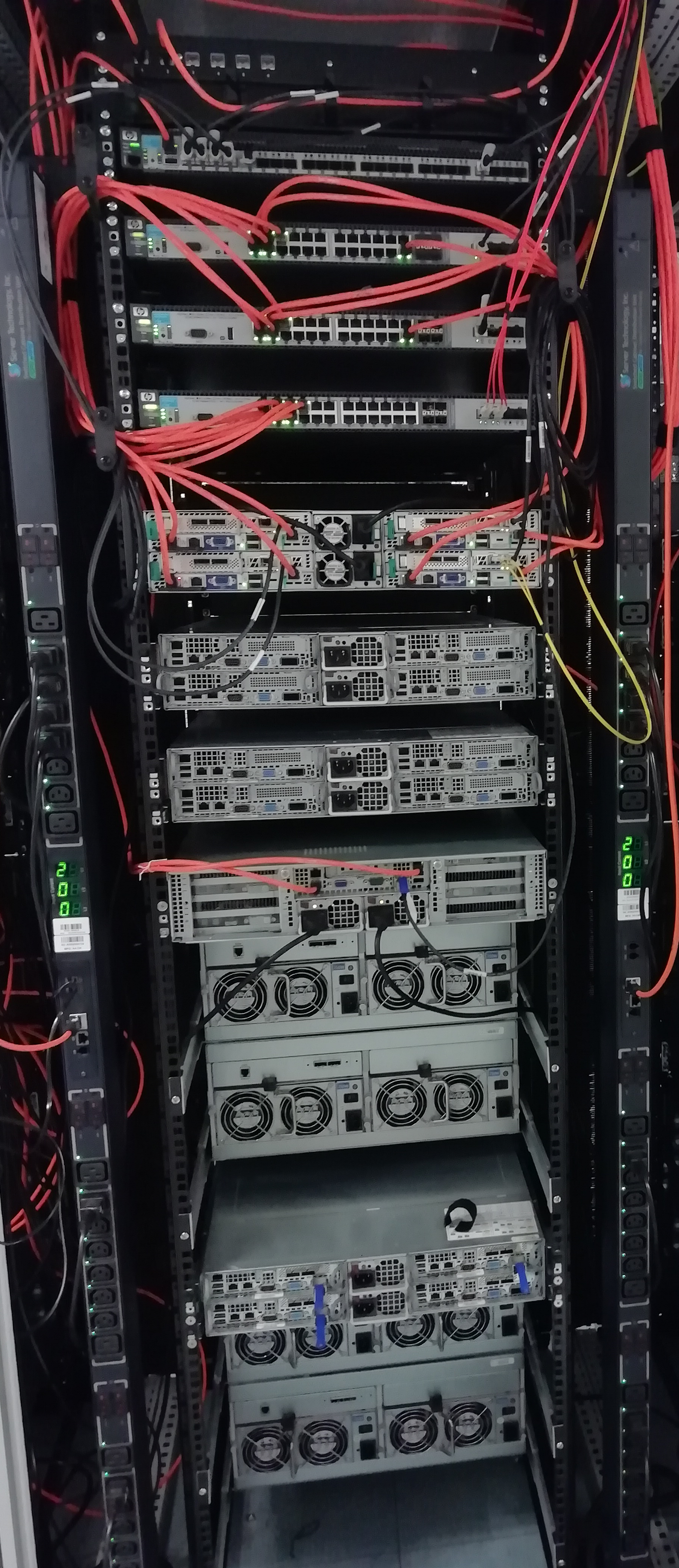}
\caption{The structure of the a single cell in the Wigner Analysis Facility.}
\label{fig:afmodule}%
\end{figure}

After several hardware tests and bandwidth optimization cycles, a dual rack-based 'cell' has been chosen as scalable unit of the Wigner AF. Such a standalone working unit is composed by compute, frontend, storage, and service elements, as illustrated in Fig.~\ref{fig:afmodule}. Each of the 8 \textit{compute chassis} includes 4 dual processor machines, totaling 1024 threads per cell. The cells process the analysis jobs submitted through a dedicated interface called VO Box~\cite{vobox}, which serves as an entry point to the AF from the global WLCG system. The jobs then passed to the HTCondor~\cite{htcondor} and HTCondor-CE~\cite{htcondorce} servers which distribute them among the connected worker nodes. 

The storage element of a cell consists of a JBOD chassis with 24 disks, controlled by the machines of the frontend chassis through XRootD~\cite{Furano_2010} and EOS~\cite{Peters:2011zz, Adde_2015} services and daemons. The collection of such File Storage Server (FST) nodes is managed by the Management Server. Each of the mentioned server machines with management services is provided with a trusted grid server certificate to ensure the seamless connection to the WLCG infrastructure.

The OS level orchestration of the machines is achieved through the Metal-As-A-Service (MAAS) data centre automation developed by Canonical~\cite{maas}. 
The co-location of compute, storage, and network nodes in the same cell serves the purpose of assuring a fast data transmission required by the analysis workflow. 
The high-speed internal communication between the nodes is ensured by HP ProCurve 6600-24XG (J9265A) switches~\cite{switch}. Utilizing the SFP+ 10 GbE ports, a high bandwidth of 10~Gbps is achieved within a cell and also between other cells.
\par

Two chassis are maintained in the first working cell uniquely for special purposes. For future developments, a compute chassis is dedicated to machines with graphic accelerator cards (GPUs), while the machines of another compute chassis serve management roles. 

\section{Computing capacity and network connectivity}
\label{subsec:capacity}

The Analysis Facility concept was tested within the CERN's ALICE experimental framework with 4 cells (8 racks in total) at the WSCLAB, comparable to a mid-sized Tier 2 site~\cite{ALICE-PUBLIC-2021-007}. The site is located at the KFKI campus, Budapest, Hungary and it is part of the Wigner Datacenter, which is connected to the GEANT network by new devices with 100~Gbps-capable link. The total storage and computing power of the site is summarized in Table~\ref{tab:summary}.

%
%\begin{figure}[H]
%\centering
%\includegraphics[width=0.47\textwidth]{structure}
%\caption{The schematic view of the Wigner AF's services.}
%\label{fig:structure}%
%\end{figure}
%
\begin{table}[H]
\centering
\caption{The summary of the total resources of the Wigner AF}
\label{tab:summary}
\begin{tabular}{cc}
\hline
\hline
 Total Storage Size & Total Computing Resource\\
\hline
% 2 redundant MGM nodes & HTCondor and HTCondor-CE \\
32 FST node & Queues for single-core and multi-core jobs \\
$24 \times 3$ TB raw capacity per FST node & 128 worker nodes \\
Total raw capacity: $\sim$2.2 PB & 32 vCPU, 64 GB RAM, for each node \\
Usable capacity with RAID-1: $\sim$1.1 PB & 4096 logical cores in total\\
\hline
\end{tabular}
\end{table}

\subsection{Benchmarking the computing resources}

\begin{figure}[H]
\centering
\includegraphics[width=\textwidth]{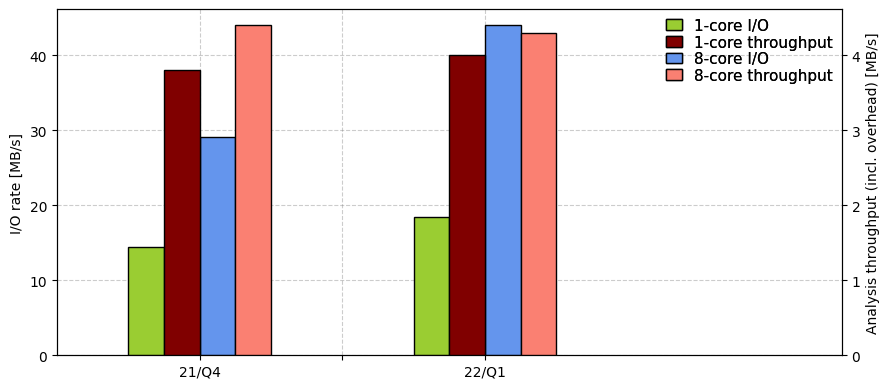}
\caption{Estimated I/O rate and analysis throughput for single- and octa-core jobs.}
\label{fig:jobs}%
\end{figure}

In parallel with the hardware installation, a set of performance and optimization tests were performed. The execution of the first, \textit{pilot} jobs occurred in late 2020, still during the hardware and software setup phase of the AF, while the production period with a high job success rate started in February, 2021. 
% In late March a second job queue was introduced for 8-core jobs (i.e. for jobs that require multi-threaded execution). It should be noted that the site was continuously tested and developed in order to fine-tune its performance, therefore the job number was fluctuating in time.\par
The performance test of the 8-rack setup with realistic analysis workload was performed in September 2021, repeated in February 2022 (see Figure \ref{fig:jobs}). The I/O rate increase, therefore the performance of the AF shows that an optimization of the analysis framework software is essential to fully utilize the capabilities of the underlying hardwares. It can be seen well, that the optimized structure results in the same I/O and +20\% analysis throughput for the single core jobs, while for octa-core ones +10\% I/O with almost the same analysis throughput.

Our tests show that the theoretical throughput (including the data I/O overhead) of the current setup has a peak at 1.1 PB/day. In average, the Wigner Analysis Facility performed $\sim 4$~MB/s analysis performance and 18~MB/s and 43~Mb/s I/O rate for the single and octa core jobs, respectively.

% On the right panel of Fig. \ref{fig:jobs} the utilization percentage of the available computing pool at a given date is shown.  
% The orange \textit{Mixed queue} bars are indicating the period when jobs with variable parallelization have been submitted to the multi-core queue, therefore the indicated rate is only a minimum value and the real utilization cannot be estimated precisely. \par
%

\section{Conclusions}

The presented Wigner Analysis Facility is one of the first instances of the throughput-specialized computing facilities that will become increasingly common among high-energy physics experiments in the future. This has been optimized for the specific task tested with the analysis framework and data provided by the CERN ALICE Experiment.  

Keeping in mind the infrastructural costs as well (electricity, High Throughput Computing hardware), the Wigner AF can provide a maintainable and scalable solution to the future computational challenges, such as gravitation wave analysis (LIGO/VIRGO) and nuclear databases within the EUPRAXIA project. This knowledge and the site is also open for other large-scale collaborations.

%The ALICE Wigner Analysis Facility is installed and operated at the Wigner Datacenter utilizing the hardware and the infrastructure used at the CERN's Tier 0 Budapest site. The main aspects during the designing phase of the AF were to keep the maintenance and upgrade costs reasonably low while optimizing the internal network between the different modules, therefore maximizing the data throughput. The first performance tests show a good throughput for both 1-core and 8-core analysis jobs, though the 8-core jobs are still to be optimized. Thanks to the modular design of the AF, the currently estimated 0.16-1.11 PB/day throughput can be further improved and upscaled with moderate effort, by introducing more modules to the infrastructure.\par

\newenvironment{acknowledgement}{\relax}{\relax}
\begin{acknowledgement}
\section{Acknowledgements}
%\input{acknowledgements_date.tex}    %%%%%%% get the lates version before submitting

% \ack
The research was supported by the Hungarian National Research, Development and Innovation Office (NKFIH) under the contract numbers OTKA K135515, and 2019-2.1.6-NEMZ\_KI-2019-00011, 2022-4.1.2-NEMZ\_KI-2022-00009, and 2022-4.1.2-NEMZ\_KI-2022-00008. The authors would like to express their gratitude to Ádám Pintér, József Kadlecsik and the technical staff of the Wigner Datacenter for the setup of the Analysis Facility hardware. We appreciate the  support of the WLCG management.

\end{acknowledgement}
\printbibliography

\end{document}